\documentclass[prl,twocolumn,showpacs]{revtex4}

\usepackage{amssymb,amsmath,bm,epsfig,graphics,subfigure}

\begin{document}

\title{Nanoparticle Shape Selection By Repulsive Interactions: \\ Metal Islands on Few Layer Graphenes}
\author{L.A. Somers$^1$, N.A. Zimbovskaya$^{2,3}$, A.T. Johnson$^1$ and E. J. Mele$^1$}
    \email{mele@physics.upenn.edu}

    \affiliation{\\ $^1$Department of Physics and Astronomy \\ University of Pennsylvania, Philadelphia PA 19104 \\ \\
     $^2$Department of Physics and Electronics, University of Puert Rico-Humacao CUH Station, Humacao 00791 Puerto Rico\\ \\
     $^3$Institute for Functional Nanomaterials \\
     University of Puerto Rico \\
     San Juan 00931 Puerto Rico}
\date{\today}

\begin{abstract}
Metal atoms adsorbed on few layer graphenes condense to form nanometer-size droplets whose growth is size limited by a
competition between the surface tension and repulsive electrostatic interactions from charge transfer between the metal droplet
and the graphene. For situations where the  work function mismatch is large and the droplet surface tension is small, a growing
droplet can be unstable to a family of shape instabilities. We observe this phenomenon for Yb deposited and annealed on few
layer graphenes and develop a theoretical model to describe it by studying the renormalization of the line tension of a two
dimensional droplet by repulsive interparticle interactions.  Our model describes the onset of shape instabilities for
nanoparticles where the growth is size-limited by a generic repulsive potential and provides a good account of the
experimentally observed structures for Yb on graphene.
\end{abstract}

\pacs{68.65.Pq, 68.70.+w, 68.37.Lp, 68.43.Hn} \maketitle

\section{I. Introduction}

\begin{figure}
\begin{center}
  \includegraphics*[angle=0,width=\columnwidth]{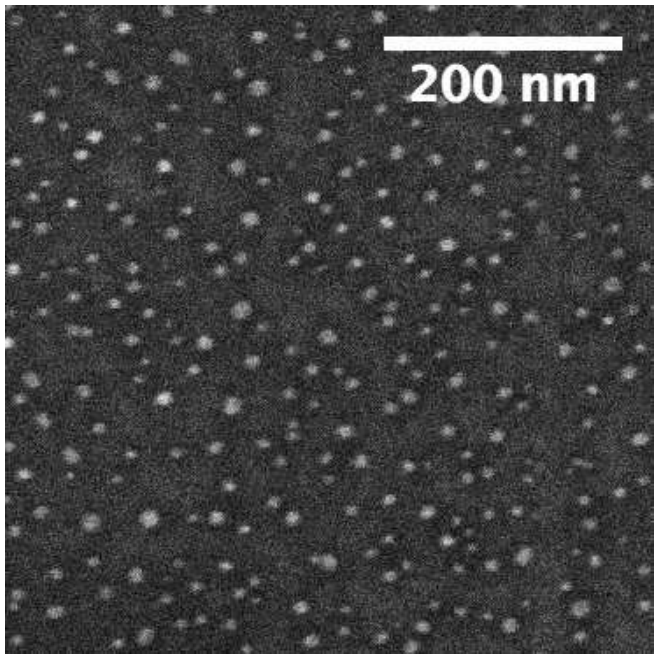}
  \caption{\label{Audrops} SEM image of Au islands  formed after annealing on four layer graphene. The Au atoms condense to form size limited
  nearly isotropic droplets whose radius is limited by the repulsion between perpendicular dipoles at the Au graphene interface. The average droplet radius
is determined
  by the microscopic width of the dipole layer, effectively the width of the few layer graphene substrate. }
\end{center}
\end{figure}

Graphite is widely used as a substrate for the synthesis of free standing metal nanoparticles due to its chemical inertness,
low diffusion barriers for adsorbed species and its compatibility with various electron microscopies. In earlier work
\cite{Luonanolett} we observed that when Au atoms are deposited on few layer graphenes (FLG)  containing $m$-layers where $1
\leq m < 20$, the mode of adatom condensation and nanoparticle growth differs sharply from that observed on thick graphite. On
few layer graphenes Au condenses to form size-limited isotropic droplets as shown in Fig. \ref{Audrops}  where the diameter is
controlled by the layer count $m$ of the graphene substrate, increasing approximately as $m^{1/3}$. This phenomenon is well
described by a model in which electrostatic dipole-dipole repulsion within a condensed island prevents the continued growth of
large droplets. The thickness dependence arises from a short range cutoff of this repulsive potential which occurs on the scale
of the dipole. For few layer graphenes, which screen poorly when the charge exchange is small, this scale is effectively the
thickness of the graphene film.

\smallskip

\medskip
\begin{figure}
\begin{center}
  \includegraphics*[angle=0,width=\columnwidth]{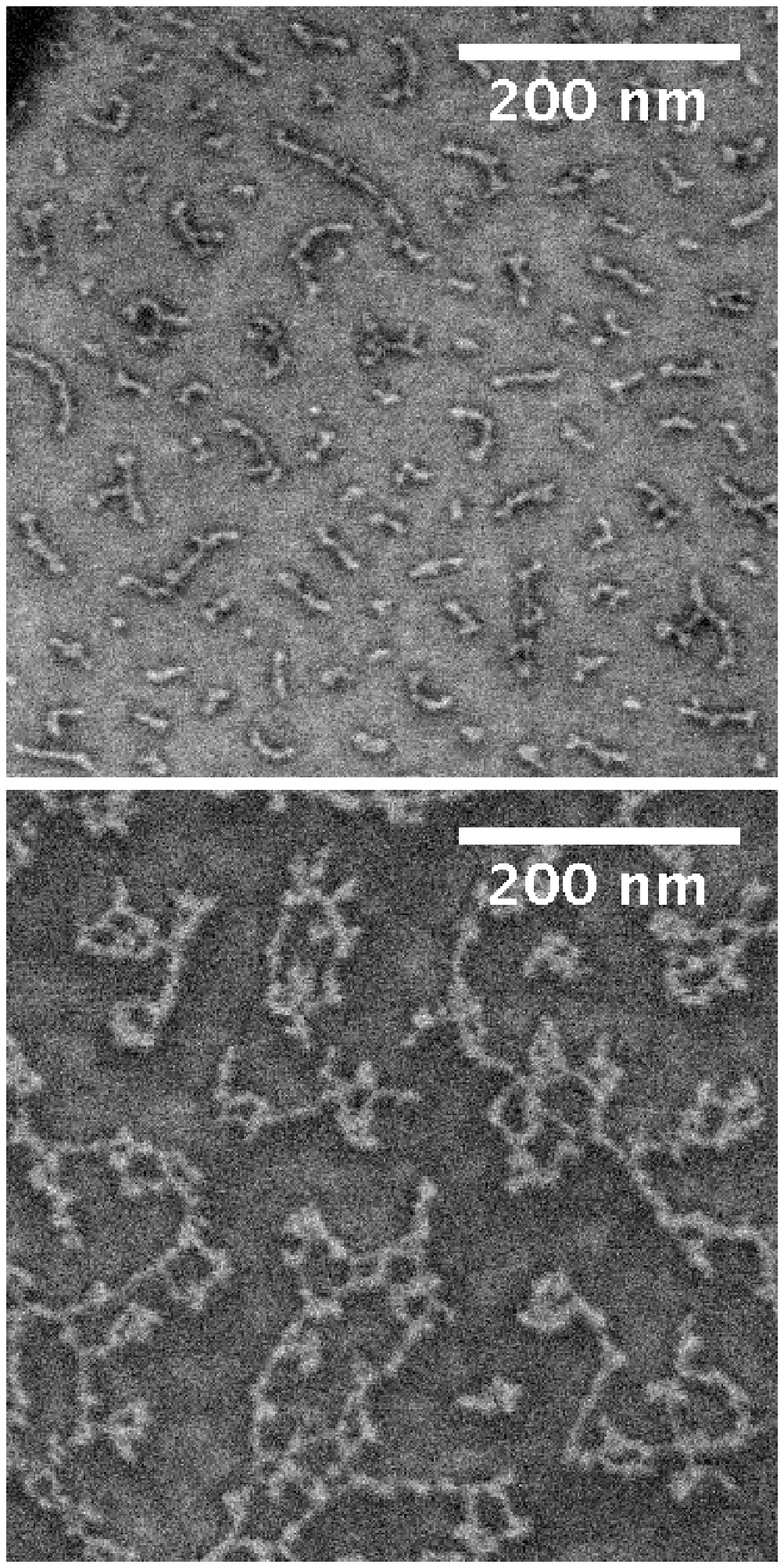}
  \caption{\label{Ybdrops} SEM image of Yb nanoparticles formed after annealing on graphene. The metal atoms condense to form
  anisotropic fibrillar structures. The top panel is for an average coverage of $0.12 \, {\rm nm}$ and the bottom
  is for an average coverage of $0.5 \, {\rm nm}$. }
\end{center}
\end{figure}

\medskip
In this paper we examine the effects of general repulsive interactions on {\it shape} selection for a condensed two dimensional
system. This work is motivated by experiments similar to \cite{Luonanolett} where Yb is adsorbed on few layer graphene films.
As shown in Fig. \ref{Ybdrops} these are observed to condense in filamentary labyrinthine structures rather than in isotropic
size-limited droplets, suggesting new physics in the surface energetics.  Indeed, Yb is distinguished by a significantly larger
workfunction mismatch to the graphene and a lower intrinsic surface tension.  These two features produce a mode-dependent
renormalization of the line tension of a growing isotropic droplet. A shape instability occurs when the line tension for a
particular surface mode of the droplet, constrained to have a constant total area, goes to zero. We develop a new formulation
describing this shape instability driven by a {\it general} repulsive potential. For application to Yb/graphene, where this
repulsive potential repulsion is provided by electrostatic dipolar interactions and can be calculated  from the work function
mismatch, we observe a sign reversal of the mode dependent surface tension for an isotropic droplet as a function of its
radius. This model correctly accounts to the observed widths of the labyrinthine structures and the occasional appearance of
threefold and fourfold vertices in the condensed filamentary patterns.

\medskip
Shape instabilities produced by electrostatic and magnetostatic dipolar interactions often arise in condensed matter problems.
For example, in two dimensions they are associated with the shapes of amphiphile domains at the air water interface
(electrostatic) \cite{McConnell} and of ferrofluid droplets (magnetostatic) \cite{Seul}. Langer {\it et al.} \cite{Langer}
examined this problem theoretically, exploiting the dipolar form of the repulsive interaction. They derived an explicit formula
for the line energy of a two dimensional droplet in terms of a double line integral over its perimeter, a result that in
principle can be used as input to an algorithm to compute the shape of a growing droplet. This approach has been refined by
Iwamoto and collaborators \cite{Iwamoto1,Iwamoto2} who replace the double line integral by an expression for the line energy
parameterized by the amplitudes of its modes of deformation, and they applied this method to study effects of dipolar
interactions both for the perpendicular geometry and for dipoles tilted with a nonzero component parallel to the tangent plane
of the droplet. Our work is similar in spirit to that of Ref. \cite{Iwamoto1}, though we  present it here in a new form that is
applicable to a general repulsive potential. For the electrostatic dipolar repulsive potential appropriate to the Yb/graphene
problem the model provides a good account of the experimental observations. Furthermore this formulation emphasizes that shape
instabilities of this type are a generic property of condensed phases of species with repulsive tails in their interaction
potentials.  Our method  can be used to access  this physics for a droplet with a general two point interaction potential in
its interior.

\medskip
Section II of this paper provides more information about our sample preparation, characterization and imaging of Yb and Au
nanoparticles formed on few-layer graphenes. Section III briefly reviews the isotropic model introduced in \cite{Luonanolett}
appropriate to size-limited circular droplets for Au on graphene. Section IV presents some useful formulas for weakly perturbed
circular droplets. Section V constructs a model for the droplet energetics, including the interaction renormalized line
tension. Section VI applied the model to Yb and Au droplets on few layer graphene and provides comparison of the model with the
experimental data. A brief discussion of the results in given in Section VII.

\section{II. Experimental}

\medskip
We prepared graphene flakes by mechanical exfoliation of kish graphite onto 300 nm silicon oxide on silicon wafers. We then
cleaned the sample by annealing at 400 C in a reducing atmosphere, 1:1 ${\rm H_2}$ and Ar. Individual flakes were identified by
color contrast in an optical microscope. Flake thickness was determined by Veeco Dimension 3100 AFM in tapping mode.

\medskip
We then thermally evaporated thin layers of metal onto the surfaces. The deposition thickness was determined by a crystal
thickness monitor. We deposited $0.22 \pm 0.1$ nm and $0.5 \pm  0.1$ nm of Yb on different chips. For Au samples, the thickness
was $0.3 \pm 0.1$ nm. Each sample was annealed to equilibrium at 600 C (Yb) or 400 C (Au) for three hours in a reducing
atmosphere, 1:1 ${\rm H_2}$ and Ar.

\medskip
We then imaged the samples in an FEI DB 235 high-resolution SEM / FIB in HRSEM mode. Even short exposures to the beam have been
found to contaminate the imaged region and reduce quality, so we were especially careful that each image was taken in a
previously unimaged area.

\medskip
Figures 1 and 2 show Au and Yb nanoparticles on single-layer and few layer graphenes. The Au particles, as noted in [1], are
nearly isotropic.The average radius of these nanoparticles increases with the thickness of the few layer graphene substrate. By
contrast Yb forms branching strands. The sample with less Yb has strands that are more regular in width.

\medskip
Figure 3 shows histograms of the distribution of Yb strand widths extracted from the images of Figure \ref{Ybdrops}. The data
at higher coverage were obtained over a wider image than shown in the lower panel of Figure \ref{Ybdrops}. We excluded only
places where the Yb was clearly branching. The mean of the distribution for 0.22 nm average coverage is $4.01 \pm 0.08$ nm, and
at a 0.5 nm coverage it increases to is $4.95 \pm 0.11$ nm. The peaks of the two distributions overlap; the increase arises
from the irregularities which are more marked on the sample with more Yb. We also measured the orientations of those Yb strands
that have a clear direction. There is no clear deviation from isotropy on the whole.

\medskip
\begin{figure}
\begin{center}
  \includegraphics*[angle=0,width=\columnwidth]{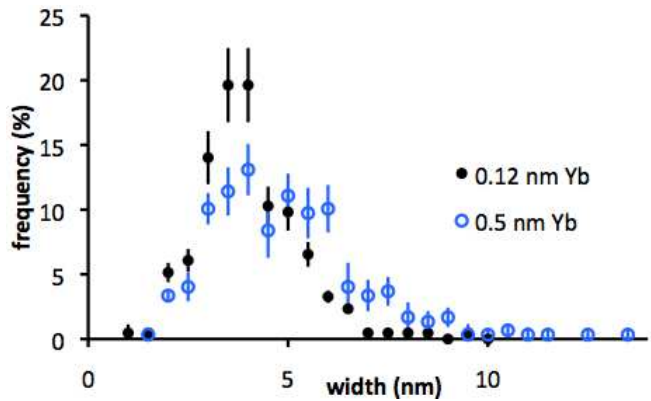}
  \caption{\label{widthdist} Width distribution of the Yb fibrils displayed in the top and bottom panels of Figure \ref{Ybdrops} for the
  average coverages shown.}
\end{center}
\end{figure}

\section{III. Review of the Isotropic Model}

Reference \cite{Luonanolett} considers the contribution to the energy of an isotropic droplet, modelled as a short circular
cylinder. The energy of a droplet of radius $R$ and height $h$ (Figure \ref{schematic})  can be expressed
\begin{eqnarray}
U = \gamma \left( \pi R^2 + 2 \pi R h \right) + e \pi R^2 h + \Gamma R^4 \nonumber
\end{eqnarray}
where $\gamma$ is the surface tension and $e$ ($<0$) is bulk cohesive energy  density. The last term results from the
dipole-dipole interactions and the scaling with $R^4$ is the essential feature that prevents the growth of large droplets. Note
that because of the long range tail of the electrostatic interaction, the last term can be regarded as a ${\rm Volume^2}$
contribution to the droplet energy. Evaluation of the coefficient $\Gamma$ in this expression requires that one regularize the
short range divergence of the $1/r^3$ dipole-dipole potential. Physically this is regularized  by the finite spatial extent of
the microscopic dipole (which we label $d$).  For graphene doped near its charge neutrality point the screening is weak and the
distance $d$ may be identified with the thickness of the graphene film \cite{Gamma}. This leads to the central result that the
droplet radius is proportional to $d^{1/3}$ which is the scaling rule identified in the experiments studying Au particles on
graphene \cite{Luonanolett}.

\medskip
The electrostatic contribution in this expression is obtained in \cite{Luonanolett}  by isolating the small momentum ($q
\rightarrow 0$) limit of the dipole-dipole interaction. By contrast, the model for the shape instability developed below will
require us to extend this into the $q \neq 0$ regime, and in fact the interactions that drive the shape instability are
obtained by integrating the interaction over all momenta.

\section{IV. Mensuration Formulae}

\begin{figure}
\begin{center}
  \includegraphics*[angle=0,width=\columnwidth]{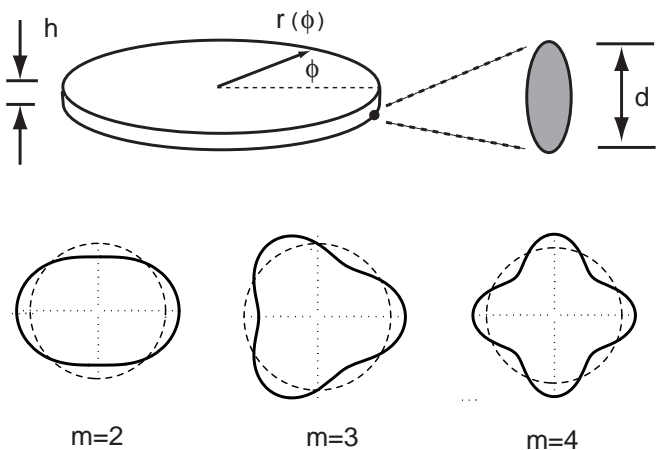}
  \caption{\label{schematic} Diagram illustrating the geometry of the deformed droplet with radius $r(\phi)$, height $h$ and
  dipole layer height $d$. The lower panel illustrates the three modes of deformation with $m=2,3,4$. }
\end{center}
\end{figure}

We are interested in perturbations of the droplet around a reference circular shape. Referring to Fig. \ref{schematic}, we
write the droplet radius as a function of the polar angle
\begin{eqnarray}
r(\phi) = r_0 + \sum_{m \neq 0} \, r_m e^{im \phi} \nonumber
\end{eqnarray}
Since $r(\phi)$ is real $r_{-m} = r_m^*$. We are considering instabilities around the circular shape and so we consider the
situation $r_m \ll r_0$.

\medskip
The differential length along the tangent line of this curve is
\begin{eqnarray}
d \ell^2 = dr^2 + (r d \phi)^2 \nonumber
\end{eqnarray}
which gives a formula for the length of the edge of the droplet
\begin{eqnarray}
{\cal L} = 2 \pi r_0 + \frac{ \pi}{r_0} \sum_m \, m^2 |r_m|^2 \nonumber
\end{eqnarray}
Note that because of the $\phi$ derivative the higher $m$ terms are more effective at increasing the arclength and thus tend to
be suppressed by a positive surface tension.

\medskip
The cross sectional area is given by a line integral over the perimeter
\begin{eqnarray}
{\cal A} = \frac{1}{2}  \oint \hat e_z \cdot (\vec r \times d \vec r)  \nonumber
\end{eqnarray}
Since $\vec r(\phi) = r(\phi) \hat e_r$ one has
\begin{eqnarray}
d \vec r = \frac{dr}{d \phi} \hat e_r + r(\phi) \hat e_\phi \nonumber
\end{eqnarray}
which gives
\begin{eqnarray}
{\cal A} = \frac{1}{2} \oint \, r(\phi)^2 \, d \phi \nonumber
\end{eqnarray}
and integrating over angles gives
\begin{eqnarray}
{\cal A} = \pi r_0^2 + \pi \sum_{m \neq 0} \, |r_m|^2 \nonumber
\end{eqnarray}
Any shape change of a droplet that preserves the particle number keeps the norm ${\cal A}$ constant (assuming fixed $h$). In
particular, if we consider an area-preserving deformation characterized by the amplitudes $r_m$ then the isotropic term has to
adjust in the manner $r_0' = r_0[r_m] = \sqrt{r_0^2[0] - \sum_m |r_m|^2} \approx r_0 - (1/2r_0) \sum_m |r_m|^2$.

\section{V. Droplet Energy}

\subsection{V.1 Line Energy}

\medskip
The effective line tension ${\cal T} = \gamma h$ giving for the line energy
\begin{eqnarray}
U_{\rm line} &=& 2 \pi \gamma h \left( r_0' + \frac{1}{2r_0'} \sum_m \, m^2 |r_m|^2  \right) \nonumber
\end{eqnarray}
where we are truncating the expansion at quadratic order in the $r_m$. The constant area constraint can be enforced by writing
\begin{eqnarray}
U_{\rm line} &=&  2 \pi \gamma h \left( r_0 + \frac{1}{2r_0} \sum_m \, (m^2 - 1) |r_m|^2  \right) \nonumber
\end{eqnarray}
Note that to quadratic order in the deformations, the $m=1$ term does not change the line energy since it describes a rigid
translation of the reference circular droplet.

\subsection{V.2 Interaction Energy}

\medskip
Two-point interactions inside the condensed droplet can modify the effective line tension at its boundary. In this section we
develop a formalism for calculating the mode-dependent renormalized line tension working in a constant-area ensemble. One can
also derive these results by working in the grand canonical ensemble, allowing for fluctuation in the total droplet area.

\medskip
We describe the interactions within the droplet by a two dimensional potential which depends on the lateral separation of two
particles $V(\vec r_1 - \vec r_2)$. In this problem we are particularly interested in the interactions between electric dipoles
mutually oriented along the interface normal with dipole density $\vec \tau$. This dipole layer produces a potential step that
equilibrates the work function mismatch between the metal and graphene $\Delta \Phi$ with $\tau = \Delta \Phi/4 \pi e$. The
electrostatic energy of an island can be expressed as a double integral over the area of the droplet
\begin{eqnarray}
U_d &=& \frac{\Delta \Phi^2}{32 \pi^2 e^2} \int  \int \, d^2 r \, d^2 r' f(\vec r)f(\vec r')g(|\vec r - \vec r'|) \nonumber
\end{eqnarray}
where the dimensionless distribution function $f(\vec r)=1$ inside the droplet and zero outside. It is useful to write this
interaction energy as a momentum integral
\begin{eqnarray}
U_d &=& \frac{\Delta \Phi^2}{32 \pi^2 e^2} \int \, \frac{d^2 q}{2 \pi} v(\vec q) |S(\vec q)|^2 \nonumber
\end{eqnarray}
where
\begin{eqnarray}
v(\vec q) = \int \frac{d^2 r}{2 \pi} \, e^{-i \vec q \cdot \vec r} g(r) \nonumber
\end{eqnarray}
and
\begin{eqnarray}
S(\vec q) &=& \int d^2 r \, e^{-i \vec q \cdot \vec r} f(\vec r) \nonumber
\end{eqnarray}
At distances large compared to the size of the dipole $d$, the interaction kernel takes the form
\begin{eqnarray}
\lim_{r \gg d} \, g(r) \rightarrow \frac{1}{r^3} \nonumber
\end{eqnarray}
which must be regularized in the near field on the scale of the dipole. A convenient form for the regularized kernel is
\begin{eqnarray}
g(r) = \frac{1}{(r^2 + d^2)^{3/2}} \nonumber
\end{eqnarray}
which gives
\begin{eqnarray}
v(q) = \frac{1}{d} \exp(- qd) = \frac{1}{d} \tilde v(\xi)  \nonumber
\end{eqnarray}
where $\xi = qd$ defines the dimensionless momentum. Note that the interaction strength scales inversely with the width of the
dipole layer $d$.

\medskip
The function $S(\vec q)$ can be parameterized in terms of the coefficients $r_m$. Consider the Fourier integral over the area
of the droplet
\begin{eqnarray}
S(\vec q) &=& \int d \phi \, r \, dr \, e^{-i \vec q \cdot \vec r} f(\vec r) \nonumber\\
 &=& \int_0^{r_0}  e^{-i \vec q \cdot \vec r}  \, r \, dr \, d \phi + \nonumber\\
  & & \,\,\,  \int_{r_0}^{r_0 + \sum_m r_m e^{im \phi}}  e^{-i \vec q \cdot \vec r} \, r \, dr  \, d \phi \nonumber
\end{eqnarray}
Here the isotropic part gives
\begin{eqnarray}
S_0(q) &=& 2 \pi \int_0^{r_0} \, J_0(qr) \,r \, dr \nonumber\\
  &=& \frac{2 \pi r_0}{q} J_1(qr_0) \nonumber
\end{eqnarray}
while the anisotropic piece can be decomposed by noting that $r_m \ll r_0$ and thus
\begin{eqnarray}
S_m = 2 \pi i^m e^{i m \phi_q} J_m(qr_0) r_0 r_m \nonumber
\end{eqnarray}
and therefore we have that
\begin{eqnarray}
S(\vec q) =\frac{2 \pi r_0}{q} J_1(qr_0) + 2 \pi \sum_m  i^m e^{i m \phi_q} J_m(qr_0) r_0 r_m \nonumber
\end{eqnarray}
Using this in our expression for the interaction energy, we find that it can be partitioned into an isotropic part
\begin{eqnarray}
U_0 = \left( \frac{\Delta \Phi^2}{8 e^2 d} \right) \int \, d \xi  \, \tilde v(\xi)  \frac{ J_1^2(\xi r_0/d)}{\xi} \,\, r_0^2 =
g_0  r_0^2  \nonumber
\end{eqnarray}
and a shape dependent piece which is given by a sum over $m$
\begin{eqnarray}
U_{{\rm shape}}  &=&  \left( \frac{\Delta \Phi^2 }{8 e^2 d} \right) \left(\frac{r_0}{d} \right)^2 \times\nonumber\\
 & & \,\,\, \sum_m  \left[ \int
\, \xi \, d\xi \, \tilde v(\xi) J_m^2(\xi r_0/d) \right] |r_m|^2  \nonumber\\
 &=& \sum_m \, g_m(r_0) |r_m|^2
 \nonumber
\end{eqnarray}
We observe that the coefficient of the $|r_m|^2$ term is always {\it positive} and thus $U_{\rm shape}$ is positive definite. A
shape instability arises from a competition between this energy and compensating changes to the isotropic interaction energy
which necessarily occur if the droplet area (equivalently the number of condensed particles) is held fixed.

\medskip
\subsection{V.3 Area Preserving Deformations}

\medskip
We wish to isolate from  the interaction energy the part that can be associated with a simple renormalization of the surface
tension (it depends only on the droplet area) and a residual part which contributes to the effective line tension (depends on
the droplet shape). To do this we write
\begin{eqnarray}
U =  U \pm \gamma_{\rm int} {\cal A}  = \tilde U + \gamma_{\rm int} {\cal A} \nonumber
\end{eqnarray}
where ${\cal A} = \pi (r_0^2 + \sum_m |r_m|^2)$.  An appropriate choice of $\gamma_{\rm int}$ removes from the energy
difference $\tilde U$ its area-dependent part and isolates the pure shape dependent energy.  $\tilde U$ can be expressed in
terms of $\gamma_{\rm int}$ and the coefficients $[g_m]$ of the quadratic terms in the interaction energy
\begin{eqnarray}
\tilde U = (g_0 - \pi \gamma_{\rm int}) r_0^2 + \sum_m \, (g_m  - \pi \gamma_{\rm int}) |r_m|^2 \ \nonumber
\end{eqnarray}
Setting $\partial \tilde U/
\partial {\cal A}=0$ eliminates the differential area dependence from $\tilde U$ and solving for $\gamma_{\rm int}$ then gives the contribution of the
interaction term to the area-dependent energy. This gives
\begin{eqnarray}
\frac{\partial \tilde U}{\partial {\cal A}} &=&   \frac{1}{2 \pi r_0} \frac{\partial \tilde U}{\partial r_0} + \sum_m
\frac{1}{2 \pi r_m} \frac{\partial \tilde U}{\partial r_m} - \gamma_{\rm int} = 0 \nonumber
\end{eqnarray}
The interaction contribution to the surface tension is
\begin{eqnarray}
\gamma_{\rm int} &=&   \frac{1}{2 \pi}\left( \frac{1}{ r_0} \frac{\partial }{\partial r_0} + \sum_m \frac{1}{ r_m}
\frac{\partial }{\partial r_m} \right) \sum_{\rm all \, n} \, g_n |r_n|^2 \nonumber
\end{eqnarray}
The residual energy $\tilde U = \sum_m \left(g_m - \pi \gamma_{\rm int} \right) |r_m|^2$ contains the shape-dependent energy
under the constraint of constant total droplet area.

\begin{figure}
\begin{center}
  \includegraphics*[angle=0,width=\columnwidth]{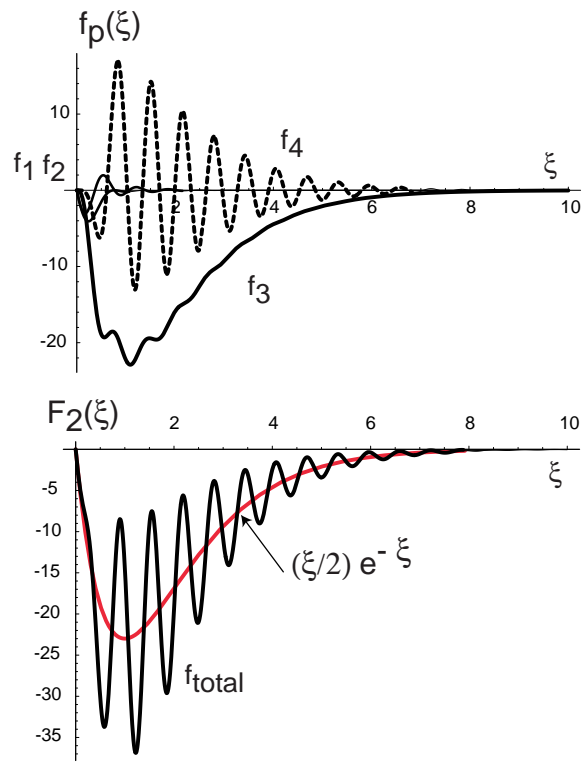}
  \caption{\label{Fandapprox} Momentum dependence of the integrand for evaluation of shape dependent interaction energy for $m=2$ plotted as a function of the
  dimensionless momentum $\xi = qd$, broken into separate contributions $f_p$ sorted by their dependence on $(r_0/d)^p$ (top panel)
  and the total (bottom panel). The thin (red) line
  in the bottom panel plots an expression discussed in the text that provides a good approximation to the integrated
  coupling strength.  The data are presented for $r_0/d=5$.}
\end{center}
\end{figure}

\medskip
In the evaluation of $\gamma_{\rm int}$ care must be taken to retain all terms that contribute to the energy $\tilde U$ at
order $|r_m|^2$, noting that the expansion coefficients $[g_m]$ are themselves function of $r_0$.  The $m$-dependent terms in
$\gamma_{\rm int} {\cal A}$ can then be organized in ascending powers of the aspect ratio $r_0/d$ in the manner
\begin{eqnarray}
\gamma_{\rm int} {\cal A}  &=&   \left( \frac{\Delta \Phi^2}{8 e^2 r_0} \right) \sum_{m \neq 0} \,\, |r_m|^2 \int \, d \xi \, \xi \, \tilde v(\xi) \times \nonumber\\
  & &   \left( \frac{r_0}{d} \right)  \frac{J_1^2(\xi
 r_0/d)}{\xi^2}  \nonumber\\
 &+&   \left(\frac{r_0}{d} \right)^2   \, J_1(\xi
 r_0/d) (\frac{J_0(\xi r_0/d) - J_2(\xi r_0/d)}{2 \xi}) \nonumber\\
 & +&   \left(\frac{r_0}{d} \right)^3   \, \left( J_m^2(\xi
 r_0/d)  + \frac{1 - J_0^2(\xi
 r_0/d)}{2} \right) \nonumber\\
 & +&  \left(\frac{r_0}{d} \right)^4  \xi J_m(\xi
 r_0/d) (\frac{J_{m-1}(\xi r_0/d) - J_{m+1}(\xi r_0/d)}{2 }) \nonumber
\end{eqnarray}
When $r_0/d$ is large, this  expression is controlled by its terms containing the highest powers of $r_0/d$. We observe that
the first term at ${\cal O} (r_0/d)^3$ exactly cancels the ``bare" contribution of the $m$-th mode to the interaction energy.
The remaining term at ${\cal O} (r_0/d)^3$ ultimately dominates the momentum integral, since the last term is a rapidly
oscillating function of $\xi$ as illustrated in Fig. \ref{Fandapprox}.  The momentum-integrated interaction is seen to be {\it
negative} indicating that for area-preserving deformations repulsive interparticle interactions destabilize the isotropic shape
at sufficiently large radius. To leading order in the small quantity $d/r_0$ we find that the interaction term simplifies to
\begin{eqnarray}
\tilde U_{\rm shape}  &=&  - \left( \frac{\Delta \Phi^2}{16 e^2 r_0} \right) \sum_{m \neq 0} \, \left(\frac{r_0}{d} \right)^3
\,\, |r_m|^2 \nonumber
\end{eqnarray}

\subsection{V.4 Renormalized Line Tension}

\medskip
Combining our results for the bare and interaction renormalized line energies, for a droplet constrained to have constant area
we have
\begin{eqnarray}
U_{\rm line} &=& \sum_{m \neq 0} \,\, \pi m^2 \gamma h \frac{|r_m|^2}{r_0} \times \nonumber\\
  & & \,\,  \left[
   1 - \frac{1}{m^2} \left(1  - \Lambda  \,
  \int \, d \xi \, \xi \, \tilde v(\xi)\, F_m(\xi, z) \, \right) \right] \nonumber\\
   &=& \sum_{m \neq 0} \,\, \pi m^2 \gamma h \frac{|r_m|^2}{r_0} \lambda_m \nonumber
\end{eqnarray}
where $z=r_0/d$ and we define the dimensionless coupling constant
\begin{eqnarray}
\Lambda =  \frac{\Delta \Phi^2}{8 \pi  e^2 \gamma h}  \nonumber
\end{eqnarray}
and the weight function for the $m$-th mode
\begin{eqnarray}
F_m(\xi,z) &\equiv&   - z^3 \frac{ 1  + J_0(\xi z)J_2(\xi z)}{2}
  \nonumber\\
    & &
     -  z^4 \frac{\xi J_m(\xi z) (J_{m-1}(\xi z) - J_{m+1}(\xi z))}{2 } \nonumber
\end{eqnarray}
Because of the rapid oscillation of the $J_m$'s  for large $r_0/d$ the  integrated coupling strength is well approximated by
retaining only the {\it constant} term in the $z^3$ coefficient; in this approximation  the momentum-integrated interaction  is
independent of $m$ and we have
\begin{eqnarray}
\lambda_m &\approx&  \left[1 - \frac{1}{m^2}  \left( 1 +  \frac{\Lambda}{2}  \left( \frac{r_0}{d} \right)^3  \right) \right]
\nonumber
\end{eqnarray}
The lower panel of Figure \ref{Fandapprox} compares the complete integrand of the momentum integral to an approximate form that
retains only the constant in the $z^3$ term. Vanishing $\lambda_m$ signifies the onset of a shape instability in the $m$-th
deformation mode. Thus the critical radius for destabilizing the $m$-th mode is
\begin{eqnarray}
\frac{r_c(m)}{d}  \approx \left( \frac{2(m^2 -1)}{\Lambda} \right)^{1/3} \nonumber
\end{eqnarray}

\section{VI. Comparison with Experiment}
\begin{figure}
\begin{center}
  \includegraphics*[angle=0,width=\columnwidth]{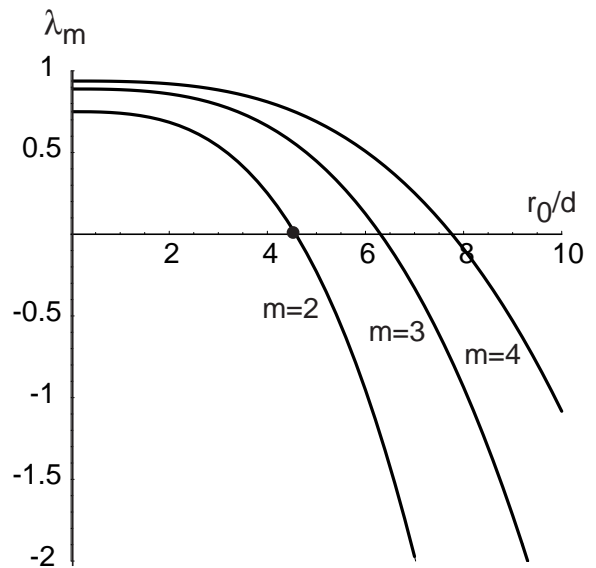}
  \caption{\label{Lambdas} Line tension renormalization coefficients $\lambda_m$ plotted as a function of $r_0/d$ for $m=2,3,4$ and for
  $\Lambda \approx .067$, appropriate to Yb/graphene.}
\end{center}
\end{figure}

For Yb on graphene we have $\Delta \Phi =2.2 \, {\rm eV}$ \cite{Ybwf}, $\gamma \approx 320 \, {\rm erg/cm^2}$ \cite{Ybgamma}
and $h \approx 10^{-7} \, {\rm cm}$ which gives $\Lambda \approx .067$. Figure \ref{Lambdas} plots the renormalization
coefficients $\lambda_m$  for this coupling strength as a function of $r_0/d$ for $m=2, 3$ and $4$, computed using the full
expression for $\gamma_{\rm int}$. The plot shows that Yb islands grows from an isotropic seed to a radius $r_0/d \sim 4.5$
where the quadrupolar $m=2$ mode becomes unstable. (The approximate expression gives $r_c(m=2)/d \approx 4.47$ and is indeed
very accurate in this regime.) Above this critical point one expects an exponential {\it growth} of the droplet along a single
axes, which one can associate with the filamentary structures observed experimentally. The width of such a filament is twice
this critical radius; for $d \approx 0.5 \, {\rm nm}$ we find that filament width  is $\sim 5 \, {\rm nm}$ in very good
correspondence with the average width $4.01 \pm .08 \, {\rm nm}$ computed from the  distribution shown in Fig. \ref{widthdist}
at $0.12 \, {\rm nm}$ average coverage. As the coverage increases the filaments fold and coarsen due mode competition between
the fibrils. The average width in the higher coverage state is thus slightly larger ($4.95 \pm 0.11 \, {\rm nm}$). One can
contrast this with the situation for Au adsorption on graphene \cite{Luonanolett} (where $\Delta \Phi \approx 0.5 \, {\rm eV}$,
$\gamma \approx 1130 \, {\rm erg/cm^2}$ \cite{Augamma}, and $\Lambda \sim \times 10^{-3}$) which gives $r_c(m=2)/d \approx 20$.
This requires growth of a circular droplet to a diameter exceeding $20 \, {\rm nm}$ which is larger than both the intrinsic
size limit imposed by the isotropic term in the dipolar energy and the largest droplet sizes observed experimentally for Au on
few layer graphenes \cite{Luonanolett}.

\medskip
In general the stability limits for the isotropic and distorted drops show different scaling with the width of the dipole layer
$d$. A size-limited circular droplet has a radius that increases  $\propto d^{1/3}$ while it is stable against shape
fluctuations below a critical radius $\propto d$. Thus in the limit of weak coupling (large $d$) a size-limited circular
droplet is stable, while for strong electrostatic coupling  with $d \sim$ interlayer spacing the droplet can undergo an
interaction-driven shape instability. Au and Yb provide examples, respectively, of this weak coupling and strong coupling
behavior.

\medskip
\medskip

\section{VII. Discussion}

\medskip
Three features of Fig. \ref{Lambdas} are noteworthy. First, the instability is driven by the terms in our expansion for the
shape energy that are quadratic in the deformation amplitudes and have the strongest (highest power) dependence on $r_0/d$.
Thus the instability requires only growth of a droplet past a critical radius for the shape change to occur.  At larger $r_0/d$
there is no mechanism that can restore a stable isotropic solution. Thus sufficiently large droplets are absolutely unstable to
this type of shape instability. Second, the renormalization coefficient shows that the crossover from the weakly renormalized
regime (small $r_0$) regime to the unstable regime (large $r_0$) occurs over a narrow size range. This also reflects the very
strong $z^3$ dependence of the dominant term in the integrated coupling strength. Thus one expects the fibrils to show a
sharply peaked width distribution, as is demonstrated experimentally in Fig. \ref{widthdist}.  Third, it is striking that the
integrated coupling strength is nearly the same for all the modes of deformation of the circular droplet. Ultimately the shape
instability is suppressed for large $m$ modes because of the $m^2$ scaling of the {\it bare} line tension rather than through
the  residual $m$ dependence in the interaction contribution. This leads to our simple scaling rule for the $m$-dependent
critical radii.

\medskip
 Finally, we note that previous continuum formulations of this problem inevitably require a finite droplet height
 to regularize the short distance singularity in the dipole-dipole potential when they are treated as {\it point} dipoles \cite{Langer,Iwamoto1}. In our treatment
this is regularized more naturally by representing the interaction potential using a nonsingular near field form controlled by
the finite size of the relevant {\it microscopic} dipoles. More generally our expression for the interaction-renormalized line
energy now can be applied to any repulsive two point potential in the droplet interior. Indeed the scaling form for $F_m$
describes the renormalization of the line tension by any nonsingular bulk interaction.

\section{Acknowledgements}

\medskip
This work was supported by the Department of Energy under Grant DE-FG02-ER45118 (EJM), by the  National Science Foundation
under PREM Grant 0353730 (NZ) and by the NSF under Grant DMR08-05136 (LS,ATJ). We thank Z. Luo and P. Nelson for helpful
discussions.


\begin{thebibliography}{11}
\bibitem{Gamma}  An explicit formula for the $d$-independent coefficient in terms of the work function mismatch is derived in \cite{Luonanolett}.
\end{thebibliography}
\end{document}